\documentclass[12pt]{article}
\usepackage{epsfig}
\usepackage{amsmath}
\usepackage{hhline}
\usepackage{amssymb}
\usepackage{times}
\usepackage{cite}

\newlength{\dinwidth}
\newlength{\dinmargin}
\begin{document}  
\newcommand{\pom}{{I\!\!P}}
\newcommand{\reg}{{I\!\!R}}
\newcommand{\slowpi}{\pi_{\mathit{slow}}}
\newcommand{\fiidiii}{F_2^{D(3)}}
\newcommand{\fiidiiiarg}{\fiidiii\,(\beta,\,Q^2,\,x)}
\newcommand{\n}{1.19\pm 0.06 (stat.) \pm0.07 (syst.)}
\newcommand{\nz}{1.30\pm 0.08 (stat.)^{+0.08}_{-0.14} (syst.)}
\newcommand{\fiidiiiful}{F_2^{D(4)}\,(\beta,\,Q^2,\,x,\,t)}
\newcommand{\fiipom}{\tilde F_2^D}
\newcommand{\ALPHA}{1.10\pm0.03 (stat.) \pm0.04 (syst.)}
\newcommand{\ALPHAZ}{1.15\pm0.04 (stat.)^{+0.04}_{-0.07} (syst.)}
\newcommand{\fiipomarg}{\fiipom\,(\beta,\,Q^2)}
\newcommand{\pomflux}{f_{\pom / p}}
\newcommand{\nxpom}{1.19\pm 0.06 (stat.) \pm0.07 (syst.)}
\newcommand {\gapprox}
   {\raisebox{-0.7ex}{$\stackrel {\textstyle>}{\sim}$}}
\newcommand {\lapprox}
   {\raisebox{-0.7ex}{$\stackrel {\textstyle<}{\sim}$}}
\def\gsim{\,\lower.25ex\hbox{$\scriptstyle\sim$}\kern-1.30ex%
\raise 0.55ex\hbox{$\scriptstyle >$}\,}
\def\lsim{\,\lower.25ex\hbox{$\scriptstyle\sim$}\kern-1.30ex%
\raise 0.55ex\hbox{$\scriptstyle <$}\,}
\newcommand{\pomfluxarg}{f_{\pom / p}\,(x_\pom)}
\newcommand{\dsf}{\mbox{$F_2^{D(3)}$}}
\newcommand{\dsfva}{\mbox{$F_2^{D(3)}(\beta,Q^2,x_{I\!\!P})$}}
\newcommand{\dsfvb}{\mbox{$F_2^{D(3)}(\beta,Q^2,x)$}}
\newcommand{\dsfpom}{$F_2^{I\!\!P}$}
\newcommand{\gap}{\stackrel{>}{\sim}}
\newcommand{\lap}{\stackrel{<}{\sim}}
\newcommand{\fem}{$F_2^{em}$}
\newcommand{\tsnmp}{$\tilde{\sigma}_{NC}(e^{\mp})$}
\newcommand{\tsnm}{$\tilde{\sigma}_{NC}(e^-)$}
\newcommand{\tsnp}{$\tilde{\sigma}_{NC}(e^+)$}
\newcommand{\st}{$\star$}
\newcommand{\sst}{$\star \star$}
\newcommand{\ssst}{$\star \star \star$}
\newcommand{\sssst}{$\star \star \star \star$}
\newcommand{\tw}{\theta_W}
\newcommand{\sw}{\sin{\theta_W}}
\newcommand{\cw}{\cos{\theta_W}}
\newcommand{\sww}{\sin^2{\theta_W}}
\newcommand{\cww}{\cos^2{\theta_W}}
\newcommand{\trm}{m_{\perp}}
\newcommand{\trp}{p_{\perp}}
\newcommand{\trmm}{m_{\perp}^2}
\newcommand{\trpp}{p_{\perp}^2}
\newcommand{\alp}{\alpha_s}

\newcommand{\alps}{\alpha_s}
\newcommand{\sqrts}{$\sqrt{s}$}
\newcommand{\LO}{$O(\alpha_s^0)$}
\newcommand{\Oa}{$O(\alpha_s)$}
\newcommand{\Oaa}{$O(\alpha_s^2)$}
\newcommand{\PT}{p_{\perp}}
\newcommand{\JPSI}{J/\psi}
\newcommand{\sh}{\hat{s}}
\newcommand{\uh}{\hat{u}}
\newcommand{\MP}{m_{J/\psi}}
\newcommand{\PO}{I\!\!P}
\newcommand{\xpom}{x_{\PO}}
\newcommand{\ttbs}{\char'134}
\newcommand{\xpomlo}{3\times10^{-4}}  
\newcommand{\xpomup}{0.05}  
\newcommand{\dgr}{^\circ}
\newcommand{\pbarnt}{\,\mbox{{\rm pb$^{-1}$}}}
\newcommand{\gev}{\,\mbox{GeV}}
\newcommand{\WBoson}{\mbox{$W$}}
\newcommand{\fbarn}{\,\mbox{{\rm fb}}}
\newcommand{\fbarnt}{\,\mbox{{\rm fb$^{-1}$}}}
%
%
\newcommand{\qsq}{\ensuremath{Q^2} }
\newcommand{\gevsq}{\ensuremath{\mathrm{GeV}^2} }
\newcommand{\et}{\ensuremath{E_T} }
\newcommand{\eq}{\ensuremath{E_T^2/Q^2} }
\newcommand{\xbj}{\ensuremath{x_{Bj}}}
\newcommand{\rap}{\ensuremath{\eta_{\rm lab}} }
\newcommand{\gp}{\ensuremath{\gamma^*}p }
\newcommand{\dsiget}{\ensuremath{{\rm d}\sigma_{ep}/{\rm d}E_t} }
\newcommand{\dsigrap}{\ensuremath{{\rm d}\sigma_{ep}/{\rm d}\eta^*} }
\newcommand{\mur}{\ensuremath{\mu_R^2}}
\def\Journal#1#2#3#4{{#1} {\bf #2} (#3) #4}
\def\NCA{\em Nuovo Cimento}
\def\NIM{\em Nucl. Instrum. Methods}
\def\NIMA{{\em Nucl. Instrum. Methods} {\bf A}}
\def\NPB{{\em Nucl. Phys.}   {\bf B}}
\def\PLB{{\em Phys. Lett.}   {\bf B}}
\def\PRL{\em Phys. Rev. Lett.}
\def\PRD{{\em Phys. Rev.}    {\bf D}}
\def\ZPC{{\em Z. Phys.}      {\bf C}}
\def\EJC{{\em Eur. Phys. J.} {\bf C}}
\def\CPC{\em Comp. Phys. Commun.}

\begin{titlepage}

\noindent
DESY 02-079  \hfill  ISSN 0418-9833 \\
June 2002


\vspace*{3cm}

\begin{center}
\begin{Large}

{\bf Measurement of Inclusive Jet Cross-Sections\\in Deep-Inelastic
  \boldmath{\em ep} Scattering at HERA}

\vspace*{1cm}

H1 Collaboration

\end{Large}
\end{center}

\vspace*{3cm}

\begin{abstract}
A measurement of inclusive jet cross-sections in deep-inelastic $ep$ scattering 
at HERA is presented based on data with an integrated luminosity of 
${\rm 21.1~pb}^{-1}$. The \mbox{measurement} is performed for photon 
virtualities \qsq between 5 and ${\rm 100~GeV}^2$, differentially in $Q^2$, 
in the jet transverse energy $E_T$, in $E_T^2/Q^2$ and in the pseudorapidity 
$\eta_{\rm lab}$. 
With the renormalization scale $\mu_R = E_T$, perturbative QCD 
calculations in next-to-leading order (NLO) give a good description of the data 
in most of the phase space. Significant discrepancies are observed only for 
jets in the proton beam direction with $E_T$ below 20~GeV and $Q^2$ below 
${\rm 20~GeV}^2$. This corresponds to the region in which NLO 
corrections are 
largest and further improvement of the calculations is thus of particular 
interest.
\end{abstract}

\vspace{1.5cm}

\begin{center}
To be submitted to Phys. Lett.
\end{center}

\end{titlepage}

%
%

\begin{flushleft}

C.~Adloff$^{33}$,              
V.~Andreev$^{24}$,             
B.~Andrieu$^{27}$,             
T.~Anthonis$^{4}$,             
A.~Astvatsatourov$^{35}$,      
A.~Babaev$^{23}$,              
J.~B\"ahr$^{35}$,              
P.~Baranov$^{24}$,             
E.~Barrelet$^{28}$,            
W.~Bartel$^{10}$
% ...

\bigskip{\it
% ...
$ ^{4}$ Inter-University Institute for High Energies ULB-VUB, Brussels;
          Universiteit Antwerpen (UIA), Antwerpen; Belgium$^{ c}$ \\
% ...
$ ^{10}$ DESY, Hamburg, Germany \\
% ...
$ ^{23}$ Institute for Theoretical and Experimental Physics,
          Moscow, Russia$^{ l}$ \\
 $ ^{24}$ Lebedev Physical Institute, Moscow, Russia$^{ e}$ \\
% ...
$ ^{27}$ LPNHE, Ecole Polytechnique, IN2P3-CNRS, Palaiseau, France \\
 $ ^{28}$ LPNHE, Universit\'{e}s Paris VI and VII, IN2P3-CNRS,
          Paris, France \\
% ...
$ ^{33}$ Fachbereich Physik, Bergische Universit\"at Gesamthochschule
          Wuppertal, Wuppertal, Germany \\
% ...
$ ^{35}$ DESY, Zeuthen, Germany \\
% ...
$ ^c$ Supported by FNRS-FWO-Vlaanderen, IISN-IIKW and IWT \\
% ...
$ ^e$ Supported by the Deutsche Forschungsgemeinschaft \\
% ...
$ ^l$ Partially Supported by Russian Foundation
      for Basic Research, grant    no. 00-15-96584 \\
}

\end{flushleft}

\newpage

\section{Introduction}

\noindent
HERA offers excellent possibilities to test predictions of quantum 
chromodynamics (QCD), the theory of the strong interaction, in 
deep-inelastic electron\footnote{
In the following, the generic name ``electron'' will be 
used for the beam and scattered lepton.
}
proton scattering (DIS). This theory has been successfully tested to a very 
high level of precision in the measurements of the proton structure function 
$F_2$~\cite{lit:h1f22,lit:zeusf2}, in which the photon virtuality $Q^2$ gives 
rise to a hard scale. QCD also predicts the production of partons with large 
transverse momenta, which fragment into hadronic jets with similar four-momenta. 
Jet observables therefore give direct access to the underlying parton dynamics. 
In addition to $\sqrt{Q^2}$, the transverse energies $E_T$ of the resulting 
jets, measured in an appropriate frame of reference, provide a natural hard 
scale for the description of the interaction within perturbative QCD.

Jets have been studied extensively at HERA and at other colliders. Recent 
measurements of (multi-)jet cross-sections at 
HERA~\cite{lit:h12jets,lit:h13jets,lit:zeus2jets,lit:zeusalphas} 
have shown that at sufficiently large values of the photon virtuality 
$Q^2 \gsim {\rm 150~GeV}^2$, calculations implementing matrix elements to  
${\cal O}(\alpha_s^2)$ in the strong coupling constant, i.e. next-to-leading 
order (NLO), are in excellent agreement with the data.
These highly successful calculations are performed in the conventional 
collinear approximation referred to as DGLAP~\cite{dglap}. Since in this 
kinematic regime of large $Q^2$,
the theoretical uncertainties are small, the measured cross-sections 
have been used to extract $\alpha_s$~\cite{lit:h12jets,lit:zeusalphas} and the 
gluon density
of the proton~\cite{lit:h12jets}. For some observables these data can be 
described even at lower values of $Q^2$, although the theoretical uncertainties
in the predictions and the QCD corrections between leading order (LO) and NLO 
predictions increase in this kinematic region.

Problems have also been 
encountered in the theoretical description of jets in DIS. 
Measurements of highly energetic jets at large 
pseudorapidities\footnote{
The pseudorapidity $\eta_{\rm lab}$ is defined as 
$\eta_{\rm lab}=-\ln\tan(\theta/2)$, with the polar angle $\theta$ being 
measured with respect to the positive $z$-axis, which is given by the proton 
beam (or forward) direction.
}
$\eta_{\rm lab}$ with transverse energies squared of the order of $Q^2$ 
(so-called forward jets~\cite{lit:bartelsetal}) made by the
H1~\cite{lit:h1forward} and ZEUS~\cite{lit:zeusforward,zeuset2q2}
collaborations show large discrepancies between data and NLO calculations at 
low values of Bjorken-$x$.
This may be due to a breakdown of the DGLAP approximation and the onset of
BFKL~\cite{bfkl} effects where, in contrast to DGLAP, the emissions of partons
are not ordered in transverse momentum.
Alternatively, a good description of the data can be obtained~\cite{jung} by
considering the partonic structure of the virtual photon, 
which is expected to be important
for $E_T^2 > Q^2$ and $Q^2$ not too large. It has been pointed 
out~\cite{lit:kramerpoetter}, however, that there is some correspondence between 
higher order effects and resolved virtual photon contributions. Further 
discrepancies between NLO calculations and data were revealed by H1 measurements
of DIS dijet production for $Q^2 < {\rm 11~GeV}^2$~\cite{lit:h1lowq2} and for 
$Q^2 < {\rm 10~GeV}^2$ 
with
invariant dijet masses $M_{jj} < {\rm 25~GeV}$ or 
mean dijet $E_T < {\rm 20~GeV}$~\cite{lit:h12jets}. In these regions of phase 
space the NLO corrections to the LO cross-sections are large, such that
next-to-next-to-leading order (NNLO) contributions are also likely to
be important.

While the H1 measurements of dijet~\cite{lit:h12jets} and three-jet 
production~\cite{lit:h13jets} mentioned above cover low and high $Q^2$ values, 
inclusive jet production in DIS 
has only been measured previously at high values 
of $Q^2$ above 150~GeV$^2$~\cite{lit:h12jets}. This paper presents precision 
measurements of inclusive jet cross-sections in the Breit frame\footnote{
The Breit frame is defined by $2x{\vec p} + {\vec q} = 0$, where $x$ is the 
Bjorken scaling variable, and ${\vec p}$ and ${\vec q}$ are the proton and 
the virtual photon momenta, respectively.
} at low $Q^2$ values (${\rm 5} < Q^2 < {\rm 100~GeV}^2$). 
Here, ``inclusive'' means 
that in every selected event every jet that passes the experimental cuts 
contributes to the measured cross-section. The advantages of investigating 
QCD with studies of inclusive jets compared with multi-jets are two-fold. 
First, the accessible phase space is extended. Second, phase space regions in
which fixed order calculations are infrared-sensitive are naturally avoided.
In dijet analyses such regions are usually suppressed by imposing asymmetric 
cuts on the transverse jet energies~\cite{lit:frixione}.

\section{\label{section:detector}Experimental apparatus}

\noindent
A detailed description of the H1 apparatus can be found 
elsewhere~\cite{lit:h1,lit:h12}. Here, only the components of the detector 
which are relevant for the measurement of low $Q^2$ jet cross-sections are 
introduced.

The scattered electron is measured with a calorimeter made of lead and 
scintillating fibres (SpaCal)~\cite{spacal}. The SpaCal covers polar angles
from ${\rm 153^{\circ}}$ to ${\rm 177.5^{\circ}}$ for collisions at the 
nominal interaction point. It is divided into an electromagnetic section with 
28 radiation lengths and a hadronic section and has a total depth of 2 
interaction lengths. The electromagnetic section of the SpaCal has an energy 
resolution of 
$\sigma_E/E \approx {\rm 0.07}/\sqrt{E [\mathrm{GeV}]} \oplus {\rm 0.01}$ 
for electrons~\cite{lit:spacalresol}.
In addition, a backward planar drift chamber in front of the SpaCal with an 
angular acceptance of ${\rm 151^{\circ}} < \theta < {\rm 177.5^{\circ}}$
serves to suppress photoproduction background, where a high energy hadron 
fakes an electron. 

The hadronic final state is measured using the SpaCal and the liquid argon 
(LAr) calorimeter~\cite{lar} together with the tracking chambers. The LAr 
calorimeter provides full azimuthal coverage over the polar angle range 
${\rm 4^{\circ}} < \theta < {\rm 154^{\circ}}$ with a depth ranging between 
4.5 and 8 interaction lengths, depending on the polar angle. Test beam 
measurements of the LAr calorimeter modules showed an energy resolution of 
$\sigma_E/E \approx {\rm 0.50}/\sqrt{E [\mathrm{GeV}]} \oplus {\rm 0.02}$ for 
charged pions after software energy reweighting~\cite{lit:larresol} and of 
$\sigma_E/E \approx {\rm 0.12}/\sqrt{E [\mathrm{GeV}]} \oplus {\rm 0.01}$ for 
electrons~\cite{lit:larresol2}.
 
The calorimeters are surrounded by a superconducting solenoid which provides
a uniform magnetic field of 
1.15~T parallel to the beam axis
in the region of the central tracking detectors. Charged particles 
are measured in this 
central tracking area which consists of drift and proportional chambers. The 
drift chambers cover a range in polar angle from ${\rm 15^{\circ}}$ to 
${\rm 165^{\circ}}$. The tracking chambers also serve for the determination 
of the event vertex.

\section{\label{section:theory}QCD calculations and Monte Carlo models}

\noindent
For the NLO calculation of jet observables at the parton level the DISENT 
computer program~\cite{lit:disent} was used. Several other 
programs~\cite{lit:jetvip,lit:mepjet,lit:disaster} are known to give 
comparable 
results~\cite{lit:nlocompare}. All of these programs calculate 
the direct 
photon induced contributions to the cross-section. 
Only the JetViP 
program~\cite{lit:jetvip} provides the additional possibility to calculate a 
cross-section consisting of both direct and resolved photon contributions. 
While implementing the concept of the resolved hadronic substructure of the 
photon is straightforward for photoproduction, in the DIS case conceptual 
difficulties are encountered~\cite{jetvipproblem} which lead to ambiguous 
results. Therefore we decided not to use JetViP for comparisons with 
the present data.

In the DISENT program the square of the renormalization scale, $\mu_R^2$, 
can be set to any linear combination of the two hard scales in the event, 
$E_T^2$ and $Q^2$. Here, $E_T$ is the transverse energy of the jet in the Breit 
frame in the case of events in which only one jet satisfies the selection 
criteria. In the case of two or more selected jets, $E_T$ was set to the mean 
transverse energy of the two hardest jets. For most of the comparisons with 
data $\mu_R = E_T$ was chosen, since $E_T^2 > Q^2$ in almost all of the phase 
space considered. However, the effects of choosing $\mu_R = \sqrt{Q^2}$ are 
also discussed in section~\ref{section:results}. The factorization scale was 
always set to $\sqrt{Q^2}$. In the kinematic region considered in this analysis
the cross-section predictions are stable within a few percent even for large 
variations of the factorization scale~\cite{mohrdis}. The parton density 
functions (PDF) of the proton were taken from the CTEQ5M (CTEQ5L) 
parameterization~\cite{cteq} for the calculation of the NLO (LO) cross-sections. 
For NLO, the corresponding value of $\alpha_s(M_Z)$ is 0.118. The number of 
active flavours was chosen to be $n_f = 5$ unless $\mu_R^2$ was less than 
25~GeV$^2$ in which case $n_f = 4$ was used in order to simulate the threshold 
for the onset of beauty production via the boson gluon fusion process.

In contrast to fixed order QCD calculations, which can only 
predict
the 
partonic final state of an event, the implementation of phenomenological QCD 
models in Monte Carlo (MC) generators allows the details of the hadronic final 
state to be 
simulated. 
These generators typically implement ${\cal O}(\alpha_s)$
matrix elements and account for 
the effects of higher orders using different 
QCD-inspired mechanisms. LEPTO~\cite{lit:lepto} and RAPGAP~\cite{lit:rapgap}
are so-called ME+PS Monte Carlo programs which combine the ${\cal O}(\alpha_s)$
matrix elements with parton showers which take into account the leading 
logarithms in $Q^2$ to all orders. As an alternative approach, the 
\mbox{ARIADNE} MC~\cite{lit:ariadne} simulates parton cascades as a chain of 
independently radiating colour dipoles according to~\cite{lit:cdm}. Higher 
order QED corrections, which can influence the event topology and the size of 
the cross-sections, are implemented in the program HERACLES~\cite{lit:heracles}
which is interfaced to RAPGAP. An interface to LEPTO and ARIADNE is provided by 
the DJANGO~\cite{lit:django} program. In the present analysis the MC generators
ARIADNE and LEPTO were used to estimate the hadronization corrections 
which were
applied 
to the NLO QCD calculations. RAPGAP and the DJANGO-interfaced ARIADNE program 
('DJANGO/ARIADNE') were used to correct the data for detector and higher order 
QED effects (see section~\ref{correction}). For all MC generators used in the 
analysis, the hadronization of the final partonic system as well as the 
particle decays were modelled with the Lund colour string model~\cite{lit:lund}
as implemented in JETSET~\cite{lit:jetset}.

\section{\label{section:measurement}Measurement}

\subsection{\label{section:selection}Event and jet selection}

\noindent
The analysed data were collected with the H1 detector in the years 1996 and 
1997. In this running period HERA collided 27.5~GeV positrons with 820~GeV 
protons. The integrated luminosity as measured using the bremsstrahlung 
process $ep \rightarrow ep\gamma$ amounts to 21.1~${\rm pb^{-1}}$.

The events are triggered by demanding a localized energy deposition in the 
SpaCal and loose track requirements. The trigger efficiencies are close to 
100~$\%$. DIS events are selected by identifying the scattered electron in the
SpaCal. More precisely, the electron is defined as the highest energy cluster 
in the SpaCal with an energy of at least 10~GeV. It must be detected at a polar
angle $\theta > {\rm 156^{\circ}}$ to guarantee the full reconstruction of the 
particle within the SpaCal acceptance.

The kinematic variables $x$, $Q^2$ and $y$ are determined using electron 
information only, according to $Q^2 = 4E_eE'_{e}\cos^2(\theta_e/2)$, 
$y = 1-(E'_e/E_e)\sin^2(\theta_e/2)$ and $x = Q^2 / (ys)$, where $E_e$, $E'_e$
and $\theta_e$ are the electron beam energy, the energy of the scattered 
electron and the electron scattering angle, respectively. The latter two 
quantities are measured using 
the SpaCal. 
The center-of-mass 
energy squared $s$ of the proton-electron scattering is computed from the 
beam energies. 

The kinematic range considered in this analysis is further constrained by the
conditions ${\rm 5} < Q^2 < {\rm 100~GeV}^2$ and ${\rm 0.2} < y < {\rm 0.6}$. 
The latter condition also leads to a reduction of photoproduction background
and of events in which the incoming electron radiates a high-energy photon. 

The hadronic final state objects of an event are reconstructed using 
information from energy deposits in the calorimeters and from tracks in the 
inner detectors. A cut $\sum_j (E_j - P_{z,j}) > {\rm 45~GeV}$ further 
suppresses radiative events. Here the sum runs over the objects of the hadronic
final state and the scattered lepton. An additional cut 
$\sum_j (E_j - P_{z,j}) < {\rm 65~GeV}$ reduces non-$ep$ background. For all 
events the reconstructed $z$-coordinate of the event vertex is required to be 
within $\pm {\rm 35~cm}$ of its nominal position which further reduces non-$ep$ 
background to a negligible level.

Jets are defined using the inclusive $k_\perp$ cluster 
algorithm~\cite{lit:ktclus} in the Breit frame in which the photon and the 
parton collide head-on. This ensures that the jet transverse energies 
are closely related to those of the partons emerging from the hard scattering.
Jets are finally selected by requiring their transverse energy to be larger 
than 5~GeV. In addition, the jets have to be well contained within the polar 
angle acceptance of the LAr calorimeter. Therefore, a cut on the pseudorapidity
of the jets ${\rm -1} < \eta_{\rm lab} < {\rm 2.8}$ is applied.

\subsection{\label{correction}Correction procedure}

\noindent
All data distributions shown in this paper are corrected for the effects of 
limited detector acceptance and resolution and for higher-order QED effects 
using a bin-to-bin correction method. Following studies of migrations made with
simulated events, the bin sizes were chosen in order to ensure that this method 
can be applied. In all bins, the stability\footnote{
The stability (purity) is defined as the number of jets which are
both generated and reconstructed in an analysis bin, divided by the 
total number of 
jets that are generated (reconstructed) in that bin.
} 
and purity exceed 40$\%$. The distributions of kinematic variables 
and
jet observables in the data are sufficiently well described by the QCD 
Monte Carlo models used in the correction procedure. RAPGAP describes the data 
very well for transverse energies above 10~GeV but lies below the data for 
$E_T < {\rm 10}$~GeV. The DJANGO/ARIADNE program gives a good description for 
$E_T < {\rm 10}$~GeV, but is too high for higher values of the transverse 
energy.

The correction factor is defined as the ratio of the number of events from two 
different MC event samples generated for the same phase space. The first sample
has no QED corrections and is not subjected to the detector simulation. The 
second sample includes QED corrections to the leptonic vertex of the interaction
and is subjected to a full simulation of the H1 detector and reconstruction
procedure. The 
correction factors were calculated from both the RAPGAP and the DJANGO/ARIADNE 
programs. The final correction was taken to be the average of the two results.
The corrections factors thus derived are typically between 0.9 and 1.1. 
Values down to 0.7 and up to 1.3 are reached
in some
exceptional 
analysis bins,
for 
example for $E_T^2/Q^2 < {\rm 2}$ and $E_T^2/Q^2 > {\rm 50}$ for jets in the 
forward region ${\rm 1.5} < \eta_{\rm lab} < {\rm 2.8}$.
Although the shapes and
normalizations of the predicted cross-sections differ significantly between 
RAGPAP and DJANGO/ARIADNE, the difference between the correction factors 
derived with the two programs is usually as small as 5 to 10$\%$. Half of this
difference is assigned as the systematic error in the correction procedure.

In order to compare the data, which are corrected for instrumental and 
radiative effects, to the NLO QCD calculations, hadronization corrections are 
applied to the latter. They were estimated using the predictions of the Monte 
Carlo programs LEPTO and ARIADNE for the ratio of the hadron level to the 
parton level cross-sections. The average of the two predictions is taken as the
correction. The effects of hadronization reduce the NLO cross-sections by 
factors that vary from 0.85 to 0.95 with increasing jet transverse energy or 
$Q^2$~\cite{meinedis}. The uncertainty in the hadronization correction is taken
to be half the difference between the predictions of the two programs.

The treatment of systematic uncertainties is similar to that 
in~\cite{lit:h12jets,lit:h13jets}. The dominant contribution to the total 
systematic error stems from the uncertainty in the hadronic energy scale of the 
LAr calorimeter, which is taken to be 3$\%$ for this analysis. This uncertainty
leads to an uncertainty in the measured cross-sections of 10 to 15$\%$. All 
other sources of systematic uncertainties, such as the method used for the 
reconstruction of kinematic variables, the determination of the energy and 
polar angle of the scattered electron and
the model dependence of the correction 
procedure, are small by comparison.

\section{\label{section:results}Results}

\begin{figure}[p]
\begin{center}
\epsfig{file=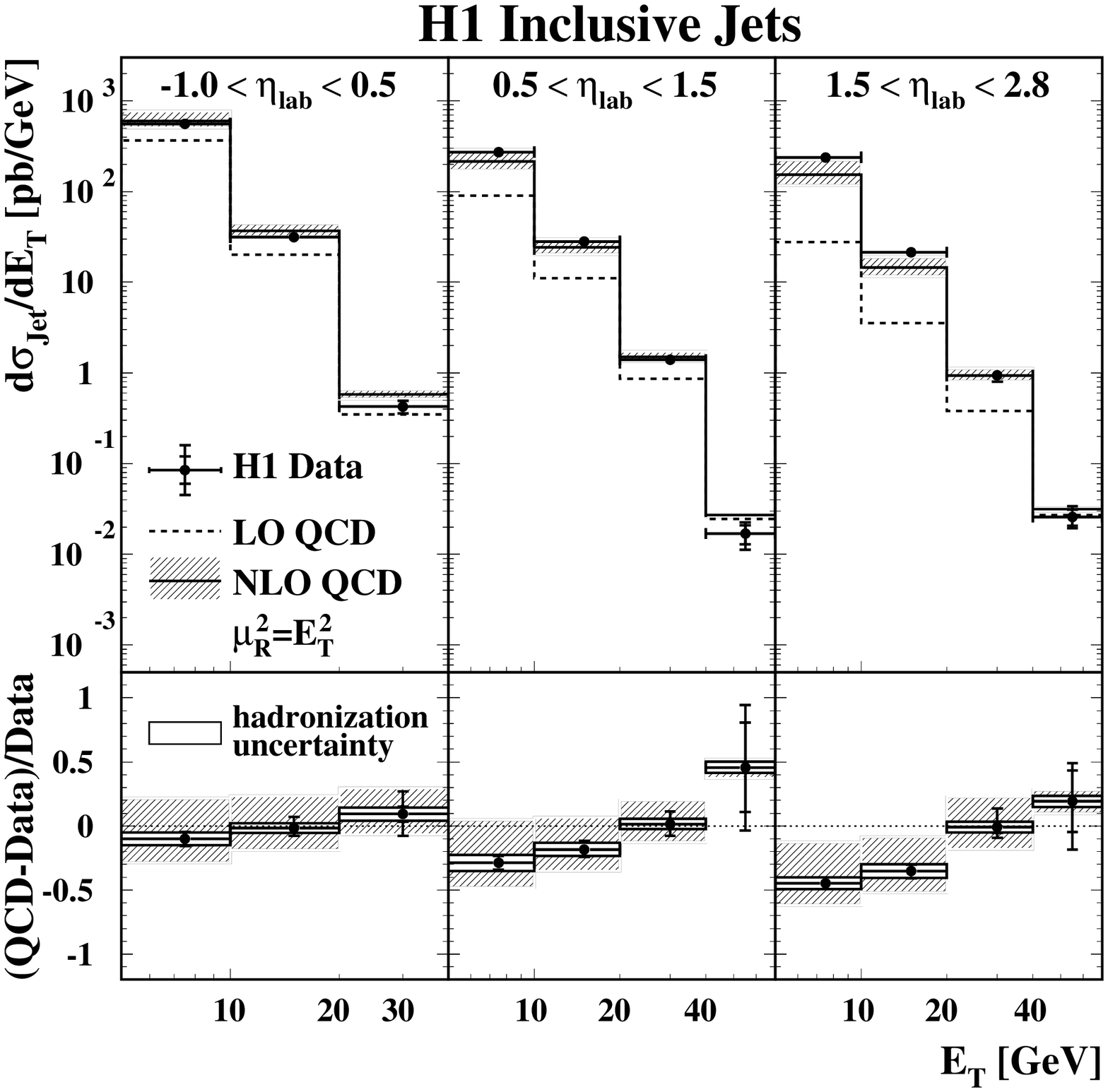, width=\textwidth}
\caption{Inclusive jet cross-sections d$\sigma_{\rm Jet}$/d$E_T$ in different 
ranges of $\eta_{\rm lab}$, integrated over the region
$5 < Q^2 < 100 \ {\rm GeV^2}$ and $0.2 < y < 0.6$. 
The data are shown as 
points with error bars which include statistical and systematic errors. In the
top part the data are compared to DISENT NLO 
QCD calculations using the CTEQ5M parton distribution 
functions (solid line) and to 
DISENT LO calculations using CTEQ5L (dashed line).
The
renormalization scale is set to 
$\mu_R=E_T$ and no hadronization corrections are 
applied to these predictions. The hatched band around the NLO prediction stems
from variations of the renormalization scale
by factors of $\pm 2$. In the bottom part the ratio
(QCD-Data)/Data is shown. Here ``QCD'' denotes NLO QCD corrected for 
hadronization effects. The inner white band shows the hadronization uncertainty.
The hatched outer area contains in addition the renormalization scale 
uncertainty added linearly. A dotted line at 0 serves to guide the eye.}
\label{fig:et_eta}
\end{center}
\end{figure}

\begin{figure}[p]
\begin{center}
\epsfig{file=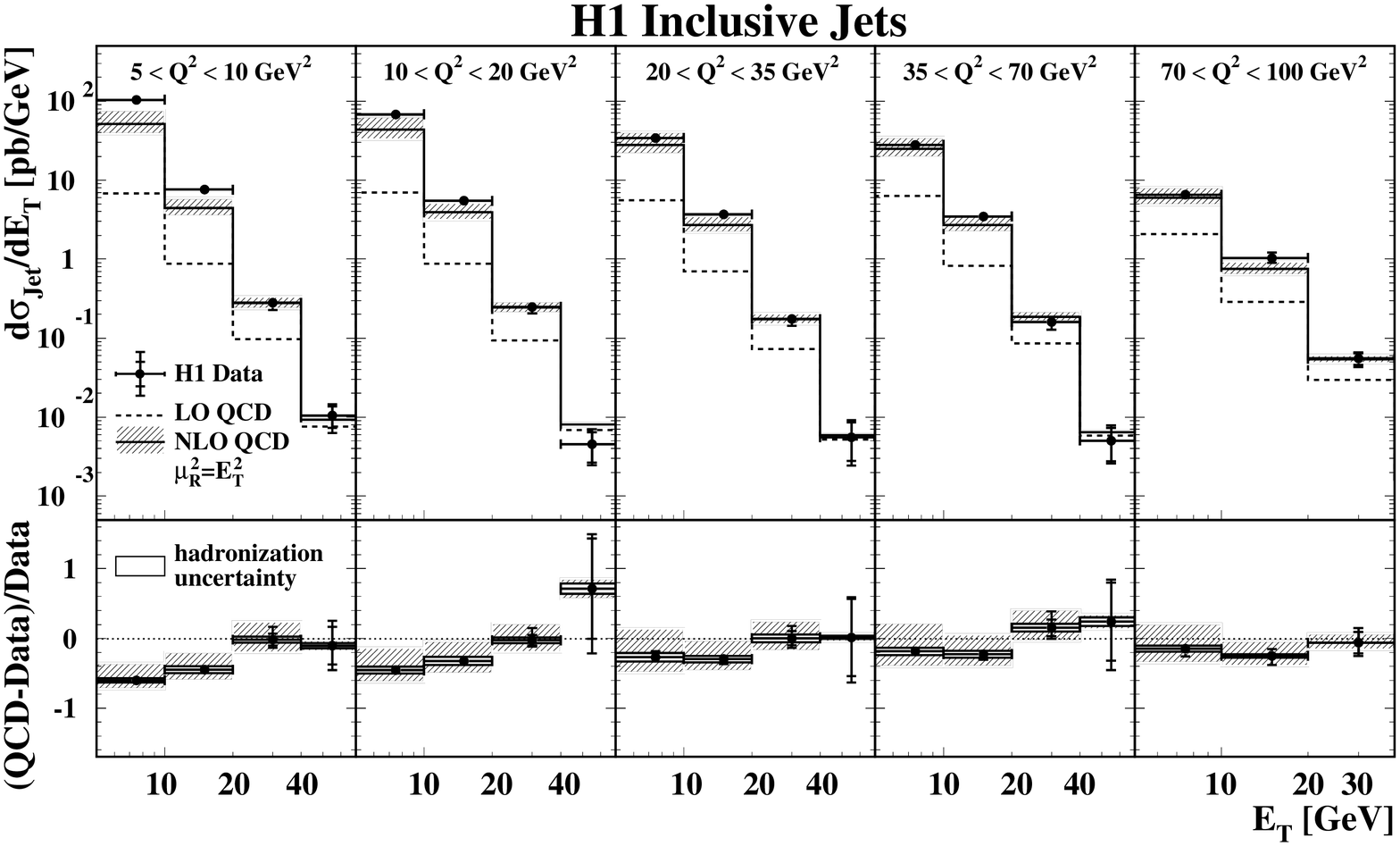, width=\textwidth}
\caption{Inclusive jet cross-sections d$\sigma_{\rm Jet}$/d$E_T$ for the forward 
region ${\rm 1.5} < \eta_{\rm lab} < {\rm 2.8}$ in different ranges of $Q^2$. 
See the caption of figure~\ref{fig:et_eta} for further explanation.}
\label{fig:et_q2}
\end{center}
\end{figure}

\begin{figure}[p]
\begin{center}
\epsfig{file=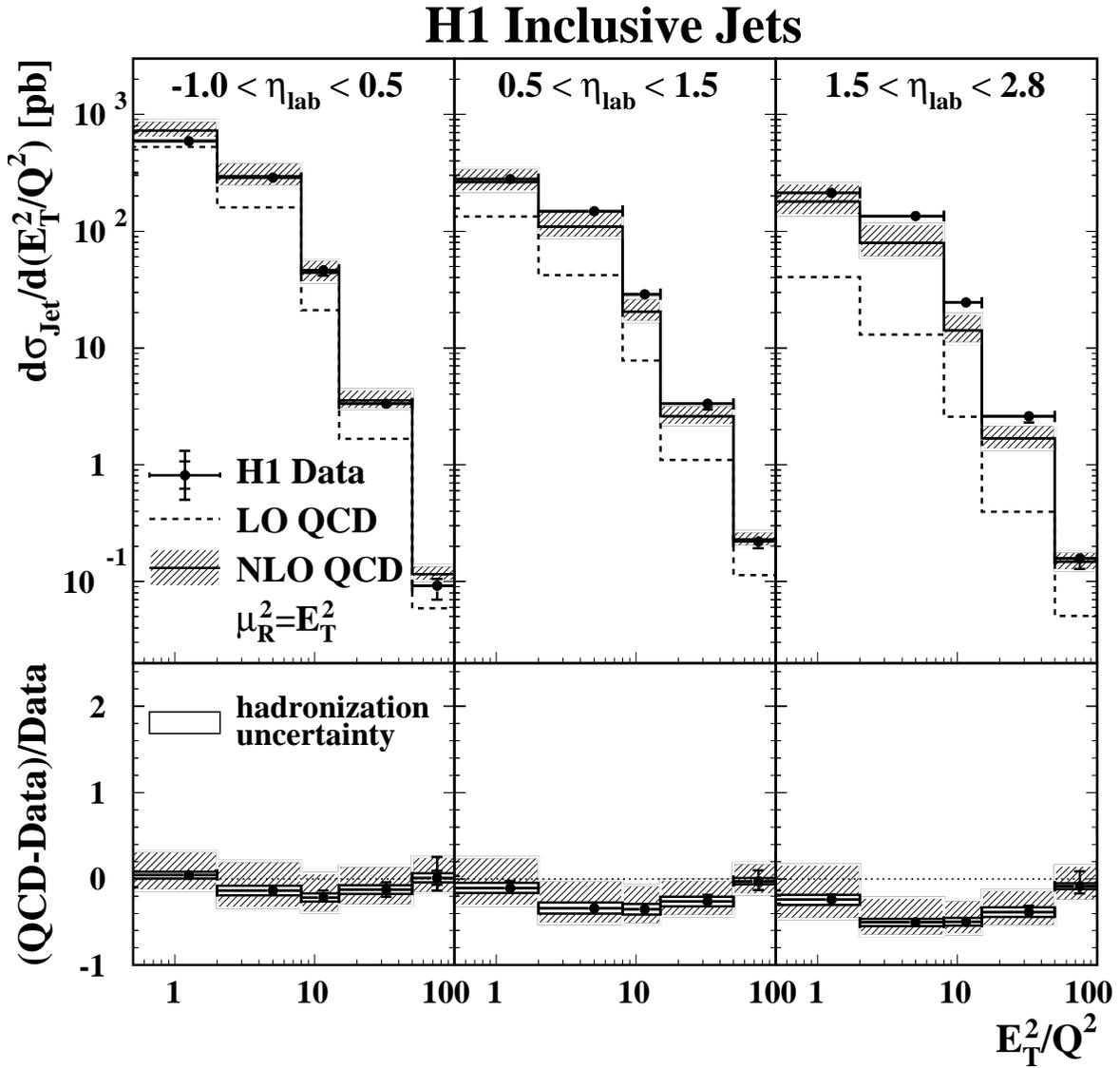, width=\textwidth}
\caption{Inclusive jet cross-sections d$\sigma_{\rm Jet}$/d($E_T^2/Q^2$) in 
different ranges of $\eta_{\rm lab}$. 
See the caption of figure~\ref{fig:et_eta} for further explanation. 
}
\label{fig:et2q2_eta}
\end{center}
\end{figure}

\begin{figure}[p]
\begin{center}
\epsfig{file=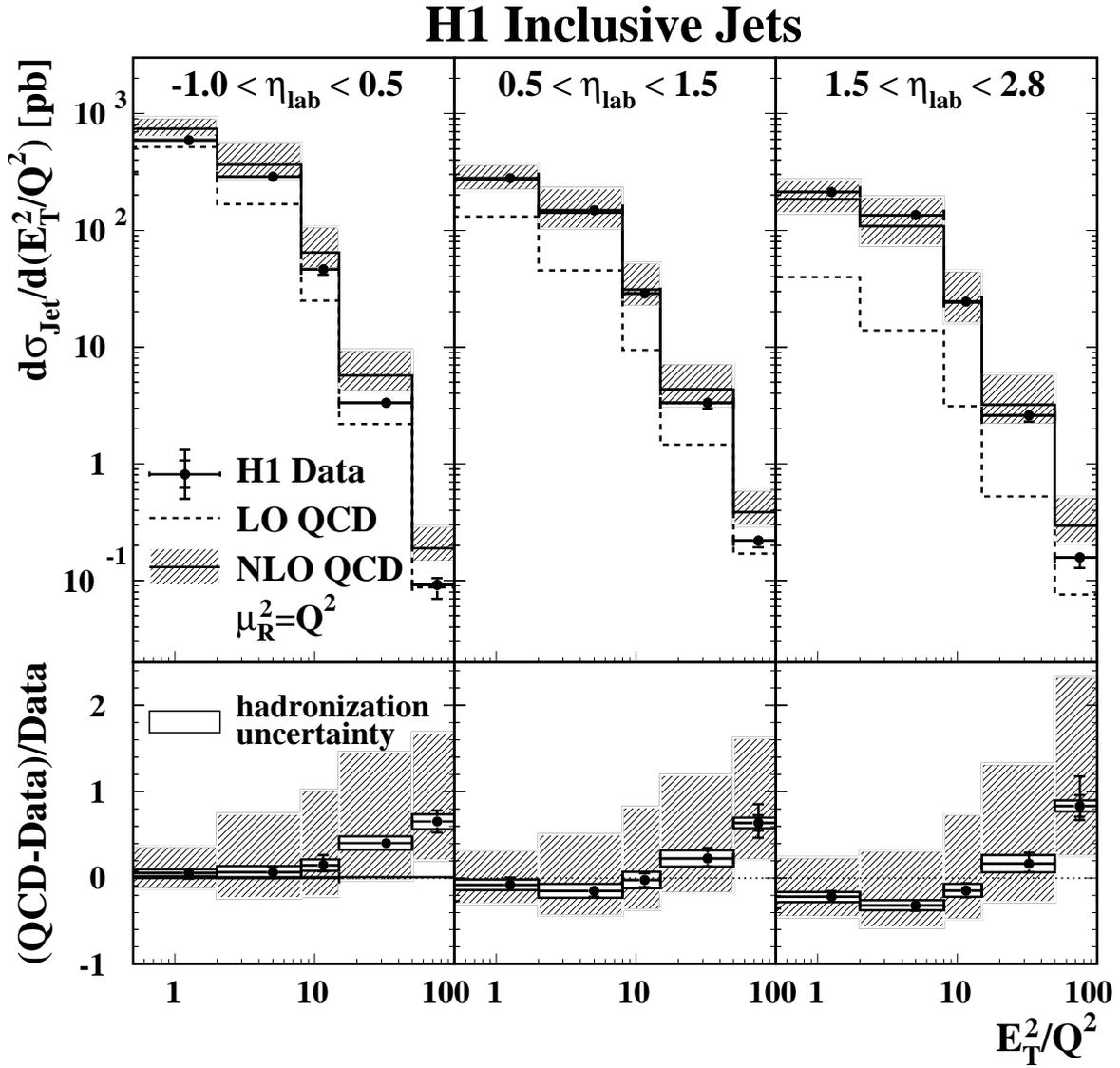, width=\textwidth}
\caption{Inclusive jet cross-sections d$\sigma_{\rm Jet}$/d($E_T^2/Q^2$) in 
different ranges of $\eta_{\rm lab}$. Here, the renormalization scale 
$\mu_R = \sqrt{Q^2}$ is used. 
See the caption of figure~\ref{fig:et_eta} for further explanation.
}
\label{fig:et2q2_eta_q2}
\end{center}
\end{figure}

\noindent
In this section the results of the inclusive jet measurement described in 
section~\ref{section:measurement} are compared with LO and NLO QCD calculations
as explained in section~\ref{section:theory}. The data are presented in 
figures~\ref{fig:et_eta} to~\ref{fig:et2q2_eta_q2}. In the top parts of the 
figures, the data, corrected for detector and radiative effects, are compared
with the results of the LO and NLO QCD calculations without hadronization 
corrections. The uncertainty in the NLO predictions, which was estimated by 
varying the renormalization scale $\mu_R$ by a factor of 
$\pm 2$, 
is also 
shown. In the bottom part of the figures, the consistency of the NLO QCD
prediction with the data is studied in more detail by plotting the relative 
difference (QCD-Data)/Data. Here ``QCD'' denotes the NLO prediction corrected 
for hadronization effects as discussed in section~\ref{correction}. Both the 
uncertainty due to the hadronization correction and the effect of the variation
of the renormalization scale $\mu_R$ are indicated. All measured 
cross-sections 
and their errors
are given in tables~\ref{table1},~\ref{table2} and~\ref{table3}, 
together with
the average Bjorken-$x$ 
and $Q^2$ values of the measurements, taken from the data.

Figure~\ref{fig:et_eta} shows the inclusive jet cross-section as a function of 
the transverse jet energy $E_T$ in different regions of the pseudorapidity 
$\eta_{\rm lab}$: in the backward region ${\rm -1} < \eta_{\rm lab} < {\rm 0.5}$, the 
central region ${\rm 0.5} < \eta_{\rm lab} < {\rm 1.5}$ and the forward (proton 
direction) region ${\rm 1.5} < \eta_{\rm lab} < {\rm 2.8}$. The measured 
cross-sections, which extend over four orders of magnitude, are compared to the
QCD calculations using a renormalization scale $\mu_R = E_T$.

While there is good agreement between 
the data and the NLO QCD calculation in the 
backward region for all $E_T$ values, discrepancies are observed for more 
forward jets with low $E_T$. At the lowest $E_T$, 5~$< E_T <$~20~GeV, for
$\eta_{\rm lab} > {\rm 1.5}$, the assumed renormalization scale uncertainty does
not cover the large difference between data and QCD calculation.
Figure~\ref{fig:et_eta} shows that the discrepancies are accompanied by large 
corrections between LO and NLO calculations. The NLO predictions are up to a 
factor 5 larger than the LO ones. Furthermore, the effects of a variation of 
the renormalization scale, which are indicated by the hatched band, are largest
for these low $E_T$ jets in the forward region.


In figure~\ref{fig:et_q2}, the jet $E_T$ distribution in the forward region 
only is presented in five different intervals of $Q^2$. Whereas for 
$Q^2 > {\rm 20~GeV}^2$ the data can be described by the NLO 
predictions even for transverse energies below 20~GeV, for 
$Q^2 < {\rm 20~GeV}^2$ the NLO calculation falls short of the data by a factor 
of up to two for $E_T < {\rm 20~GeV}$. As in fig.~\ref{fig:et_eta} these 
discrepancies between data and predictions are accompanied by large NLO/LO 
corrections.

One can conclude that the perturbative NLO QCD calculations work reasonably 
well in most of the rapidity range, even in the forward region 
$\eta_{\rm lab} > {\rm 1.5}$, as long as both $E_T$ and $Q^2$ are not too small.


In order to study the interplay of the two possible scales in DIS jet events, 
figure~\ref{fig:et2q2_eta} shows the inclusive jet cross-section as a function 
of the ratio $E_T^2/Q^2$ for the different regions of $\eta_{\rm lab}$. The data 
are well described by the NLO calculations over the full $E_T^2/Q^2$ range only
in the backward region ${\rm -1} < \eta_{\rm lab} < {\rm 0.5}$. For the 
central and even more for the forward pseudorapidities, there is a discrepancy 
between data and calculation for the medium range of this ratio, i.e. 
${\rm 2} < E_T^2/Q^2 < {\rm 50}$. This region is dominated by small values of 
$E_T^2$ and $Q^2$. For large and small values of $E_T^2/Q^2$, which are 
correlated with large values of either $E_T^2$ or $Q^2$, the NLO calculation is
in agreement with the data. 
As in the case of d$\sigma_{\rm Jet}$/d$E_T$ the discrepancies occur in
regions where the NLO/LO
corrections are very large (up to a factor of 6 in the forward region). 


Switching to  $\mu_R = \sqrt{Q^2}$ (figure~\ref{fig:et2q2_eta_q2}) changes the 
situation. In this case, large discrepancies between data and NLO calculation 
are observed only for large values of $E_T^2/Q^2 > {\rm 50}$ (and hence low
values of $Q^2$) irrespective of the $\eta_{\rm lab}$ range. This suggests that 
when $Q^2$ is much smaller than $E_T^2$, the variable $Q^2$ provides too soft a
scale and is irrelevant for the description of the hard scattering process. In 
contrast to the case $\mu_R = E_T$, these discrepancies do not occur in the 
analysis intervals in which the NLO/LO corrections are largest. Although the 
data and the predictions are compatible at lower values of $E_T^2/Q^2$, it 
should be noted that the choice of $\mu_R = \sqrt{Q^2}$ leads to much larger 
scale uncertainties than is the case with $\mu_R = E_T$. 

\section{Summary}\label{section:summary}

\noindent
Inclusive jet cross-sections have been measured for values of $Q^2$ between 
5 and 100~GeV$^2$ with the H1 detector at HERA. Jets were selected using the 
inclusive $k_{\perp}$ algorithm in the Breit frame and were required to have a 
minimum transverse energy of 5~GeV. QCD calculations up to second order of the
strong coupling constant $\alpha_s$, with $E_T$ or $\sqrt{Q^2}$ as 
renormalization scale, were tested against the data.

With $E_T$ as renormalization scale, the data on d$\sigma_{\rm Jet}$/d$E_T$ are 
well described by the NLO QCD calculations for jets of all transverse energies
in the backward region and in all $\eta_{\rm lab}$ regions for high transverse 
jet energies $E_T > {\rm 20~GeV}$, 
even though the NLO cross-sections can be more than 3 times larger than the
LO cross-sections.
The only region in which the
NLO QCD predictions fail to 
describe the data is in the forward region when both the
transverse jet energy and $Q^2$ are relatively small. In this region, 
discrepancies of up to a factor of two are observed, not considering the scale 
uncertainties. 
The regions in which the data cannot be described by the QCD 
calculations are characterized by large NLO/LO corrections and by a strong 
dependence on the renormalization scale. It is worth noting that the theoretical
uncertainties, which are dominated by the effects of a variation of the 
renormalization scale, are usually much larger than the experimental errors.

Similarly, the distributions of $E_T^2/Q^2$ are described by the QCD 
calculations using $\mu_R = E_T$ everywhere except for the range 
${\rm 2} < E_T^2/Q^2 < {\rm 50}$ for jets in the forward and central
regions. Again, the
discrepancies between data and calculation are accompanied by large NLO/LO 
corrections. For the choice $\mu_R = \sqrt{Q^2}$ the agreement between data and
predictions is in general good. However, a very strong dependence of the QCD 
calculations on the renormalization scale is found which questions whether they
can be considered as predictive. For $\mu_R = \sqrt{Q^2}$ and 
$E_T^2/Q^2 > {\rm 50}$ the calculations fail to describe the data for all jet 
pseudorapidities, showing that $Q^2$ is not an appropriate scale choice if 
$Q^2 \ll E_T^2$.

The correlation of large NLO/LO corrections and high sensitivity to 
renormalization scale variations with poor agreement between data and QCD 
predictions in the case of $\mu_R = E_T$ strongly suggests that the inclusion of 
higher order (e.g. NNLO) terms in the QCD calculations is necessary in order to 
describe the data. 
Since in most analysed intervals the experimental errors are much smaller than 
the theoretical uncertainties, it is evident that a deeper 
understanding of low 
$Q^2$ DIS jet production will require theoretical progress.
 
\section*{Acknowledgements}
\noindent
We would like to thank M. Seymour 
for many helpful discussions.
We are grateful to the HERA machine group whose outstanding
efforts have made and continue to make this experiment possible. 
We thank
the engineers and technicians for their work in constructing and now
maintaining the H1 detector, our funding agencies for 
financial support, the
DESY technical staff for continual assistance, 
and the DESY directorate for the
hospitality which they extend to the non DESY 
members of the collaboration.


\clearpage

\clearpage

\clearpage

\clearpage

\clearpage

\begin{table}[p]
\begin{center}
\footnotesize
\begin{tabular}{|c|c|c|c|c|c|c|}
\hline
\multicolumn{7}{|c|}{d$\sigma_{\rm Jet}$/d$E_T$} \\
\hline
$\eta_{\rm lab}$ & $E_T$ & value & stat. error & syst. error  & $<x>$ & $<Q^2>$\\
range & interval [GeV]& [pb/GeV] & [pb/GeV] & [pb/GeV] & [${\rm 10^{-3}}$] & [${\rm GeV}^2$]\\
\hline
\hline
        & 5 - 10 & 561 & 4 & +33 / -35 &  & \\
\cline{2-5}
-1.0 - 0.5& 10 - 20 & 31.3 & 0.5 & +2.7 / -2.0 & 0.7 & 21 \\ 
\cline{2-5}
        & 20 - 40 & 0.43 & 0.03  & +0.06 / -0.06 & & \\ 
\hline
\hline
        & 5 - 10 & 272 & 3 & +21 / -19 & & \\ 
\cline{2-5}
0.5 - 1.5& 10 - 20 & 28.2 & 0.5  & +2.3 / -2.1 & 0.8 & 23\\ 
\cline{2-5}
        & 20 - 40 & 1.40 & 0.07  & +0.12 / -0.13 & &  \\ 
\cline{2-5}
        & 40 - 70 & 0.017 & 0.006 & +0.004 / -0.004 & &  \\ 
\hline
\hline
        & 5 - 10 & 238 & 3 & +18 / -18 & & \\ 
\cline{2-5}
1.5 - 2.8& 10 - 20 & 21.4 & 0.5 & +1.6 / -1.9 & 0.8 & 21 \\ 
\cline{2-5}
        & 20 - 40 & 0.94 & 0.06  & +0.07 / -0.13 & &  \\ 
\cline{2-5}
        & 40 - 70 & 0.026 & 0.008 & +0.004 / -0.006 & &  \\ 
\hline

\end{tabular}
\caption{Measured jet cross-sections d$\sigma_{\rm Jet}$/d$E_T$ in 
different ranges of $\eta_{\rm lab}$ for ${\rm 5} < Q^2 < {\rm 100~GeV}^2$
and $0.2 < y < 0.6$, together
with their statistical and systematic uncertainties. 
The mean values of Bjorken-$x$ and $Q^2$ of the accepted events are also given.
}
\label{table1}
\end{center}
\end{table}

\clearpage

\begin{table}[p]
\begin{center}
\footnotesize
\begin{tabular}{|c|c|c|c|c|c|c|}
\hline
\multicolumn{7}{|c|}{d$\sigma_{\rm Jet}$/d$E_T$ for ${\rm 1.5} < \eta_{\rm lab} < {\rm 2.8}$} \\
\hline
$Q^2$ & $E_T$ & value & stat. error & syst. error & $<x>$ & $<Q^2>$\\
range [${\rm GeV}^2$]&  interval [GeV]& [pb/GeV] & [pb/GeV] & [pb/GeV] &
[${\rm 10^{-3}}$] & [${\rm GeV}^2$] \\
\hline
\hline
        & 5 - 10 & 104 & 3 & +9 / -11 & & \\ 
\cline{2-5}
5 - 10  & 10 - 20 & 7.6 & 0.4  & +0.6 / -0.6 & 0.2 & 7 \\ 
\cline{2-5}
        & 20 - 40 & 0.28 & 0.03 & +0.05 / -0.02 & & \\ 
\cline{2-5}
        & 40 - 70 & 0.010 & 0.005 & +0.003 / -0.002 & & \\ 
\hline
\hline
        & 5 - 10 & 68.1 & 1.6  & +4.5 / -6.0 & & \\ 
\cline{2-5}
10 - 20 & 10 - 20 & 5.5 & 0.2  & +0.4 / -0.4 & 0.5 & 14 \\ 
\cline{2-5}
        & 20 - 40 & 0.25 & 0.03 & +0.04 / -0.01 & & \\ 
\cline{2-5}
        & 40 - 70 & 0.005 & 0.003 & +0.001 / -0.002 & & \\ 
\hline
\hline
        & 5 - 10 & 34.3 & 1.4   & +4.1 / -1.6 & & \\ 
\cline{2-5}
20 - 35 & 10 - 20 & 3.7 & 0.2  & +0.3 / -0.4 & 0.9 & 26 \\ 
\cline{2-5}
        & 20 - 40 & 0.17 & 0.03 & +0.03 / -0.02 & & \\ 
\cline{2-5}
        & 40 - 70 & 0.01 & 0.005 & +0.001 / -0.002 & & \\ 
\hline
\hline
        & 5 - 10 & 28.0 & 0.8 & +1.9 / -2.1 & & \\ 
\cline{2-5}
35 - 70 & 10 - 20 & 3.5 & 0.2 & +0.2 / -0.4 & 1.7 & 49 \\ 
\cline{2-5}
        & 20 - 40 & 0.16 & 0.02 & +0.03 / -0.01 & & \\ 
\cline{2-5}
        & 40 - 70 & 0.005 & 0.005 & +0.001 / -0.002 & & \\ 
\hline
\hline
        & 5 - 10 & 6.5 & 0.4 & +0.5 / -0.8 & & \\ 
\cline{2-5}
70 - 100 & 10 - 20 & 1.03 & 0.08 & +0.12 / -0.06 & 3.4 & 82 \\ 
\cline{2-5}
        & 20 - 40 & 0.055 & 0.015 & +0.008 / -0.006 & & \\ 
\hline

\end{tabular}
\caption{Measured jet cross-sections d$\sigma_{\rm Jet}$/d$E_T$ 
for $0.2 < y < 0.6$ in 
different ranges of $Q^2$ for jets in the forward region 
${\rm 1.5} < \eta_{\rm lab} < {\rm 2.8}$, 
together with their statistical and 
systematic uncertainties.
The mean values of Bjorken-$x$ and $Q^2$ of the accepted events are also given.
}
\label{table2}
\end{center}
\end{table}

\clearpage

\begin{table}[p]
\begin{center}
\footnotesize
\begin{tabular}{|c|c|c|c|c|c|c|}
\hline
\multicolumn{7}{|c|}{d$\sigma_{\rm Jet}$/d($E_T^2/Q^2$)} \\
\hline
$\eta_{\rm lab}$ & $E_T^2/Q^2$ & value & stat. error & syst. error & $<x>$ & $<Q^2>$\\
range & interval & [pb] & [pb] & [pb] & [${\rm 10^{-3}}$] & [${\rm GeV}^2$] \\
\hline
\hline
        & 0.5 - 2 & 593 & 6 & +32 / -35 & 1.2 & 37 \\ 
\cline{2-7}
        & 2 - 8 & 286 & 3   & +17 / -18 & 0.4 & 14 \\ 
\cline{2-7}                     
-1.0 - 0.5 & 8 - 15 & 46.3 & 0.9 & +4.7 / -2.7 & 0.3 & 9 \\ 
\cline{2-7}
        & 15 - 50 & 3.3 & 0.1  & +0.3 / -0.3 & 0.2 & 8 \\ 
\cline{2-7}                     
        & 50 - 100 & 0.09 & 0.01 & +0.02 / -0.01 & 0.2 & 7 \\ 
\hline                          
\hline                          
        & 0.5 - 2 & 280 & 4 & +25 / -15 & 1.4 & 39 \\ 
\cline{2-7}                     
        & 2 - 8 & 149 & 2  & +9 / -12 & 0.5 & 15 \\ 
\cline{2-7}                     
0.5 - 1.5 & 8 - 15 & 28.9 & 0.7 & +2.2 / -2.5 & 0.4 & 11 \\ 
\cline{2-7}                     
        & 15 - 50 & 3.3 & 0.1 & +0.3 / -0.2 & 0.3 & 11 \\ 
\cline{2-7}                     
        & 50 - 100 & 0.22 & 0.02 & +0.03 / -0.02 & 0.3 & 9 \\ 
\hline                          
\hline                          
        & 0.5 - 2 & 213 & 5 & +18 / -13 & 1.4 & 38 \\ 
\cline{2-7}                     
        & 2 - 8 & 135 & 2 & +10 / -12 & 0.5 & 14 \\ 
\cline{2-7}                     
1.5 - 2.8 & 8 - 15 & 24.5 & 0.8 & +2.2 / -2.2 & 0.3 & 10 \\ 
\cline{2-7}                     
        & 15 - 50 & 2.6 & 0.1 & +0.3 / -0.2 & 0.3 & 9 \\ 
\cline{2-7}                     
        & 50 - 100 & 0.16 & 0.02 & +0.03 / -0.01 & 0.3 & 9 \\ 
\hline

\end{tabular}
\caption{Measured jet cross-sections 
d$\sigma_{\rm Jet}$/d$(E_T^2/Q^2)$ in different ranges of $\eta_{\rm lab}$ for 
${\rm 5} < Q^2 < {\rm 100~GeV}^2$ and $0.2 < y < 0.6$, 
together with their statistical and 
systematic uncertainties.
The mean values of Bjorken-$x$ and $Q^2$ of the accepted events are also given.
}
\label{table3}
\end{center}
\end{table}

\end{document}